\documentclass[12pt,preprint]{emulateapj}

\usepackage{graphicx,amsmath,natbib}

\slugcomment{Submitted to ApJL}

\shorttitle{Companions inside 10 AU in the HR 8799 system}
\shortauthors{Hinkley et al.}

\begin{document}

\title{Observational Constraints on Companions inside of 10 AU in the HR 8799 Planetary System}

\author{Sasha Hinkley            \altaffilmark{1,2}}
\author{John M. Carpenter    \altaffilmark{1}}   
\author{Michael J. Ireland  \altaffilmark{3}}
\author{Adam L. Kraus  \altaffilmark{4}}
\altaffiltext{1}{Department of Astronomy, California Institute of Technology, 1200 E. California Blvd, MC 249-17, Pasadena, CA 91125}
\altaffiltext{2}{Sagan Fellow}
\altaffiltext{3}{Sydney University}
\altaffiltext{4}{Hubble Fellow, Institute for Astronomy, University of Hawaii, 2680 Woodlawn Dr., Honolulu, HI 96822, USA }
    
\begin{abstract}
We report the results of Keck $L'$-band non-redundant aperture masking of HR 8799, a system with four confirmed planetary mass companions at projected orbital separations of 14 to 68 AU. We use these observations to place constraints on the presence of planets and brown dwarfs at projected orbital separations inside of 10 AU---separations out of reach to more conventional direct imaging methods.  No companions were detected at better than 99\% confidence between orbital separations of 0.8 to 10 AU. 
Assuming an age of 30 Myr and adopting the Baraffe models, we place upper limits to planetary mass companions of 80, 60, and 11 M$_{\rm Jup}$ at projected orbital separations of 0.8, 1, and 3-10 AU respectively. 
Our constraints on massive companions to HR 8799 will help clarify ongoing studies of the orbital stability of this multi-planet system, and may illuminate future work dedicated to understanding the dust-free hole interior to $\sim$6 AU.   


\end{abstract}


\keywords{instrumentation: adaptive optics---instrumentation: interferometers---planets and satellites: detection---stars: individual (HR 8799, HD 217165)---techniques: high angular resolution}


\section{Introduction}
HR 8799, a 20--150 Myr \citep{mad06} A5V star hosting several planets, presents a challenge for formation models of massive exoplanets.  Until recently, this system was known to have three $\sim$5-7 M$_{\rm Jup}$ exoplanets (assuming an age of 30 Myr), clearly identified with adaptive optics (``AO'') at separations of 24-68 AU \citep{mmb08}.  The recent identification of a fourth $\sim$7 M$_{\rm Jup}$ companion at 14 AU \citep{mzk10} reinforces that this system contains an intriguing architecture.  
Maintaining this system in a dynamically stable, non-resonant configuration for the 10$^7$-$10^8$ year age of the system has proved to valid only for models possessing a restricted range of physical parameters.   \citep[e.g.][]{gm09,fm10}. 
Furthermore, forming massive planets such as these at such wide projected orbital separations is problematic for the core-accretion model of planet formation \citep[e.g.][]{dvf09}.

The recent direct imaging observations dedicated to HR 8799 are most sensitive to planetary mass companions at projected orbital separations $\gtrsim$10 AU.  
However, obtaining a census of any planets within 10 AU is essential for a full understanding of the origin and stability of the system \citep[e.g.][]{rks09}.
Several authors have addressed the possibility that the placement of wide-separation exoplanets is caused by outward scattering from inner massive perturbers or a planetesimal disk \citep[e.g.][]{va04, vcf09}. The advantage of such models rests on the formation of the bodies just beyond the location of the ice line at a few Myr where the core accretion timescales are much shorter. In a similar vein, several works have analyzed the degree to which the placement of wide separation planets such as those in HR 8799 are caused by mutual outward scattering from the existing planets \citep[e.g.][]{cfm08}. Similarly, \citet{sm09} justify the existence of long period exoplanets as a natural by product of dynamical relaxation through planet-planet scattering. 

Further, several multi-planet exoplanetary systems appear to be ``dynamically packed'' \citep[][]{br04,rbv09}; i.e., with the placement of their mutual orbits lying close to a boundary for dynamical stability.  Hence, in systems with multiple companions, it is imperative to assess the full orbital distribution of planetary mass companions and test such hypotheses.  

The planets orbiting HR 8799 discovered to date have been imaged using angular differential imaging \citep[``ADI''---][]{lmd07,mld06,lsh10}. To obtain high contrast images of this system at even smaller separations, we have used non-redundant aperture masking, hereafter ``NRM'', using the NIRC2 infrared camera at the W. M. Keck Observatory. NRM is the only technique that can produce significant contrast levels at such small inner working angles ($\lesssim$ 300---500 mas) among existing 10m class telescopes.  We are able to place upper limits on additional massive companions to the HR 8799 system that would otherwise have been veiled by the telescope diffraction or the associated uncorrected quasi-static speckle noise \citep[see e.g.][]{hos07} using more conventional direct imaging techniques \citep{oh09}.  Our technique, which we describe in greater detail in Sections~\ref{obs} and \ref{analysis}, bolsters ongoing ground-based NRM efforts \citep{ik08,kim08,bbi10} as well as serving as a precursor for the NRM efforts to be used with the James Webb Space Telescope, hereafter ``JWST'' \citep{slt10,dhr10}.

\section{Methods \& Observations}\label{obs}
\subsection{Non Redundant Aperture Masking}
NRM \citep{bhm86,rnp88} achieves contrast ratios of $\sim$$10^2$ - $10^3$ at very small inner working angles, usually within $\sim$$\frac{1}{3}\lambda/D - 4\lambda/D$ ($\sim$ 20-300 mas for Keck $L'$-band imaging).  Applications of this technique \citep[e.g.,][]{tmd00, lmi06,mtd07,wtm08,bbi10} use AO along with an opaque mask containing several holes, constructed such that the baseline between any two holes samples a unique spatial frequency in the pupil plane. A key feature of this technique is its ability to sample the ``closure phase'' quantity between any three baselines \citep[e.g.][]{bhm86, hmt87}.  While the effective transmission of the telescope is reduced by $\sim$90-95\% due to the mask, the effectiveness  of NRM is limited primarily by the ability to calibrate the point spread function.  Hence the importance of measuring closure phases using carefully chosen calibrator stars exceeds the need for gathering additional flux. Assuming the aperture holes are small relative to the characteristic size of any atmospheric turbulence, and in the regime where the AO system is providing overall phase stability, the closure phase quantity is largely unaffected by atmospheric turbulence.  Under good conditions, this technique can achieve $L'$-band contrasts of $10^2$-$10^3$.   This technique is particularly well suited for studies of stellar multiplicity and the brown dwarf desert \citep[e.g.][]{kim08,ik08}, or detecting giant planets orbiting young stars ($\lesssim$~50-100 Myr).

\subsection{Observations}
We imaged the HR 8799 system using NRM operating at $L'$-band (3.76 $\mu$m)  on the nights of UT 2009 Aug 5 and 6 using AO and the NIRC2 infrared camera on the Keck 2 telescope at Mauna Kea.  While no seeing measurements were made, non-NRM observations obtained each night verified that the AO system was able to return apparently diffraction-limited images at $K$-band.  An observing sequence consisted of observing a star in both the upper left and lower right quadrants of the NIRC2 detector with the  Keck nine hole aperture mask (see Figure 1, left panel). At each detector position, we obtained 15 images with 20 s exposures. Nine and ten such 30-image sequences of HR 8799 were obtained  on the first and second nights, respectively. Observations of a calibrator star were obtained using a similar 30-image sequence in between each of the HR 8799 target sequences,  with the calibration star interferogram placed in the identical manner on the array.  A different calibrator star was observed in between each HR 8799 sequence. The HR 8799 and calibrator star observations, plus observing overheads, required $\sim$4 hours on each night.  
We selected these calibrator stars from the compilation of stars with stable radial velocity measurements by \citet{nmb02}, who found them to have RMS variations of $<$0.1 km s$^{-1}$ over an interval of $\sim$10 years; this criterion was meant to select against binary systems that would not be suitable for calibration. All calibrators were selected to be near HR 8799 on the sky ($\rho \la 15^\circ$) and to have similar optical $V$ magnitudes (to obtain similar AO correction) and similar or brighter near infrared $K$ magnitudes (to match or exceed the signal-to-noise in the HR 8799 masking data).  A summary of the observations is presented in Table~\ref{overviewtable}.

\begin{figure}[ht]
\center
\resizebox{.25\hsize}{!}{\includegraphics[angle=90]{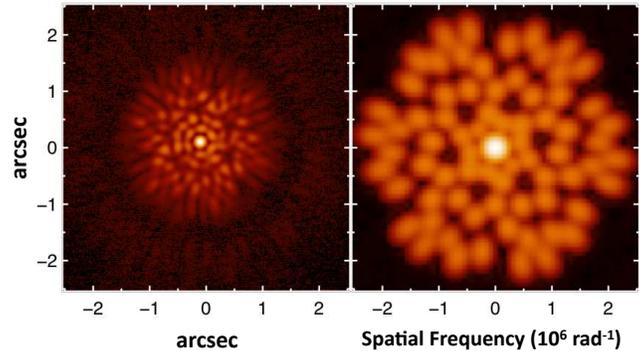}}
 \caption{{\it Left:} A Keck $L'$-band interferogram of the star HR 8799 produced with the NIRC2 nine hole non redundant aperture mask.  {\it Right:} The power spectrum corresponding to the image on the left. }
  \label{data} 
\end{figure}

\begin{deluxetable*}{lccccccll}
\tabletypesize{\scriptsize}
\tablecaption{Observing log for HR 8799 and Calibrators. 
}
\tablewidth{0pt}
\tablehead{ \colhead{Object}  & \colhead{$K$} & \colhead{Separation (deg)}\footnotemark[1] & \colhead{RA (J2000)} & \colhead{Dec. (J2000)} & \colhead{$t_{exp}$ (s)}\footnotemark[2] & \colhead{$t_{exp}$  (s)}\footnotemark[3] & \colhead{Type} & \colhead{Comment}}
\startdata                               
 HR 8799   & 5.24 &          & 23 07 28.7 & +21 08 03.3 & 270$\times$20 & 300$\times$20 & target &  \\
 HD 217813 & 5.15 &  1.12 & 23 03 05.0 & +20 55 06.9 & 30$\times$20  & 60$\times$20 & calibrator & \\
 HD 216625 & 5.73 &  3.56 & 22 54 07.4 & +19 53 31.4 & 30$\times$20  & 30$\times$20 & calibrator & \\
 HD 219172 & 6.09 &  5.98 & 23 13 48.2 & +15 22 03.5 & 30$\times$20  & 30$\times$20 & calibrator & \\ 
 HD 218133 & 5.68 &  6.70 & 23 05 31.2 & +14 27 06.0 & 30$\times$20  & 30$\times$20 & calibrator & \\ 
 HD 217165 & 6.19 & 11.53 & 22 58 29.9 & +09 49 31.9 & 60$\times$20  & 30$\times$20 & calibrator & binary (not used) \\
 HD 222033 & 5.79 & 12.08 & 23 37 06.6 & +30 40 40.9 & 30$\times$20  & 60$\times$20 & calibrator & \\
 HD 221830 & 5.30 & 12.11 & 23 35 28.9 & +31 01 01.8 &                  . . .     & 30$\times$20 & calibrator & \\
 HD 210460 & 4.47 & 14.37 & 22 10 19.0 & +19 36 58.8 & 30$\times$20  & 30$\times$20 & calibrator                     
\enddata
\label{overviewtable}
\end{deluxetable*}
\footnotetext[1]{Angular separation between the calibrator stars and the HR 8799 target, measured in degrees.}
\footnotetext[2]{Total exposure time: number of exposures multiplied by the exposure time for each image for UT 2009 Aug 5.}
\footnotetext[3]{Total exposure time for UT 2009 Aug 6.}

\section{Analysis}\label{analysis}
To use the closure phase quantity to search for planetary mass companions, we follow the analysis outlined in \citet{kim08} and \citet{ik08}. We briefly summarize the procedure here. The data are initially flatfielded, sky subtracted, aligned, and corrected for cosmic rays.   
The bispectrum, the complex triple product of visibilities defined by the three baselines formed from any three subapertures, is then calculated. The phase of this complex quantity is the closure phase.

As discussed in \citet{kim08}, the calibration procedure centers around calculating the closure phase for calibrator stars. The calibrated object closure-phase is found by subtracting a weighted average of the closure-phase for the calibrator stars.  For the analysis in this paper, which is motivated by the search for point sources, the squared visibilities were not used as they were noisier than the closure-phases. Eight separate calibrators were used, as detailed in Table~\ref{overviewtable}. Each calibrator was calibrated against all other calibrators to search for resolved calibrators. HD 217165 was found to be a binary star of separation 125 mas and 2.6 magnitudes contrast during this process, and was removed as a calibrator. Each target data set was then calibrated with each calibrator observed on the same night. The Root Mean Square (RMS) calibrated closure-phase was found for each of these target-calibrator pairs, and all calibrations that resulted in an RMS closure phase more than 1.5 times the minimum for each target over all calibrators were assigned a weight of zero. In practice, this meant that each target data set was calibrated by an average of the two or three calibrator data sets obtained closest in time. The causes of the deviations from zero closure-phases (of $\la$2 degrees RMS) in the calibrator stars and the time-dependence of calibration fidelity are not fully understood, but we have evidence that dispersion in the $L'$-band due to water-vapor is one cause. The lack of perfect calibration is still the dominant source of closure-phase noise in this analysis, however the final contrast limits are within 1 magnitude of being photon-limited.

\section{Results }
In this section, we search for companions in the the HR 8799 dataset by identifying any non-zero closure phases, to be elaborated below.  No companions were detected interior to 14 AU, the location of HR 8799e.  We thus use these data to place limits on companion masses with orbital radius. 
  
\subsection{Detection Limits} 


To find the contrast limits of our data, we first added a calibration error in quadrature to uncertainties estimated directly from the scatter in the data so that the chi-squared for a null model with no companion was equal to the number of measured closure-phases.
 We then scaled the chi-squared value by the ratio of the total number of linearly independent closure phases to the total number of closure phases (28/84) for the purpose of standard least squares fitting, which assumes independent measured quantities. We have found this technique to be nearly equivalent but much simpler than the detailed estimation of covariance matrices as detailed in \citet{kim08}.  We then simulated $10^4$ data sets with closure-phase standard deviations that matched these modified uncertainties. For each simulation, we found the best contrast at each value of separation (searching over position angle) and assigned the 99\% confidence limit to the contrast where 99\% of the simulations had no best fit companion above this limit. When fitting to the actual data using a grid search, we also found no tentative solutions above this limit, therefore we are reporting a null result.

\begin{figure*}[ht]
\center
\resizebox{1.00\hsize}{!}{\includegraphics{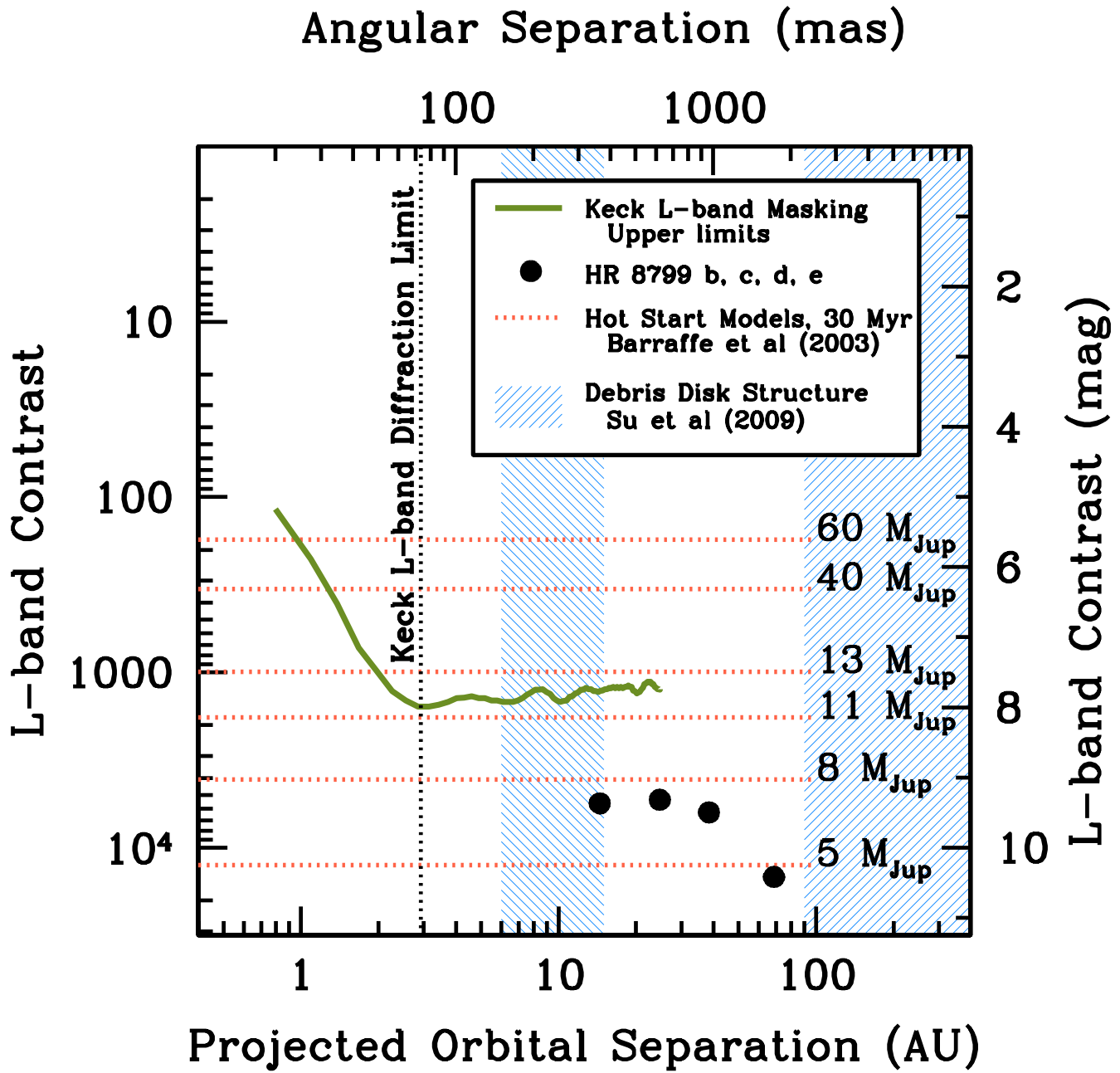}}
 \caption{Detection limits for the two combined Keck nights of non-redundant aperture masking data on the HR 8799 system.  The green line shows the minimum brightness required for a 99\% confidence detection in our HR 8799 data.  The dotted orange lines show selected corresponding mass limits as described in \cite{bcb03} assuming an age of 30 Myr.  Also shown is the theoretical diffraction limit at $L'$-band on a 10m telescope (vertical dotted line), as well as the $L'$-band brightnesses of the four known companions to HR 8799 \citep{mmb08,mzk10}. The blue regions represent  the debris disk structures defined by \citet{srs09}.}
  \label{contrasts} 
\end{figure*}

Figure~\ref{contrasts} shows our $L'$-band detection limits achieved with the two nights of Keck NRM data.   In Figure~\ref{contrasts}, we show the brightnesses and positions for the four planetary mass companions discussed in \citet{mmb08, mzk10}.  
The HR 8799 ``d'' and ``e'' companions are within the angular range probed by our NRM observations.  Both components are approximately $\Delta L'$$\sim$9.3 mag fainter than the star, while our observations achieve $\Delta L'$$\sim$7.78 and $\Delta L'$$\sim$7.74 at these orbital separations and are unable to detect these companions.   
However, we are able to probe much closer in to the host star than these separations.  We achieve our best contrast of 7.99 mags at the $L'$-band diffraction limit of the Keck telescopes, corresponding to a 3 AU projected orbital separation for HR 8799. This orbital separation is comparable to the 2-4 AU ice line boundary for a mid-A star \citep{kk08} in the 3-5 Myr timeframe for formation of gaseous giant planets. 

To convert the contrast limits to constraints on companion masses, we used the models of \citet{bcb03} along with the Hipparcos distance of 39.4 pc \citep{v07}. These models were chosen to maintain consistency with \citet{mmb08,mzk10} as well as to provide ease of comparison with other studies. We assume an age of 30 Myr, which is consistent with the 20-150 Myr age assigned to this system by \citet{mad06} as well as the two age regimes (30$^{+20}_{-10}$ and 60$^{+100}_{-30}$ Myr) discussed in \citet{mzk10}.  
\citep{mad06} assigned an age of 20-150 Myr to the HR 8799 system.  In addition, there is evidence that the system may be a member of the Columba moving group, and hence has an age of 30 Myr \citep[][R.~Doyon, B.~Zuckerman, private communication 2010]{tqm08}. \citet{mab10}, however, suggest the system may be as old as 1 Gyr.  For consistency with \citet{mmb08,mzk10}, we adopt an age of 30 Myr. 

Table~\ref{contrasttable} summarizes the sensitivity limits to companion masses for an assumed age of 30 Myr.  At a separation of 3 AU, we place an upper limit of 11 M$_{\rm Jup}$ to any companion, with a comparable sensitivity outwards to 24 AU.  Interior to 3 AU, we can place upper limits to additional companions of 13 M$_{\rm Jup}$ and 60 M$_{\rm Jup}$ at 2 AU and 1 AU, respectively.  Our measurements are also able to probe to sub-AU levels in the system placing an upper limit mass of $\sim$80 M$_{\rm Jup}$ at 0.8 AU.  
For an age of 100 Myr, the masses of the currently known exoplanets would be estimated at 9-14 M$_{\rm Jup}$, and our sensitivity at 1, 2, and 3 AU shift to 80, 30, and 22 M$_{\rm Jup}$, respectively.

\section{Constraints on Inner Material: No massive perturbers}
As discussed in the Introduction, planetary scattering is one possible model to explain the presence of four massive planets at large orbital radii in the HR 8799 system. An object interior to, and more massive than, the four 5-7 M$_{\rm Jup}$ planets would likely be necessary to scatter the objects to their current placement \citep{cfm08}.  Our observations place the most stringent constraints to date on any massive perturber interior to the orbit of HR 8799 ``e.''  The mass of any perturber must be less than $\sim$11 M$_{\rm Jup}$ between 3 and 10 AU.  It is possible, of course, that a perturber more massive than our limit was in fact present at 3 AU, but has since moved significantly inward of 3 AU or even was scattered into the star.  

The dynamical stability of this four planet system has been discussed at length in other works \citep{gm09, mrs10, fm10}. 
Stability arguments almost certainly require that the planets are both in a resonant configuration and all have masses $\la$10\,M$_{\rm Jup}$. Indeed, \citet{mzk10} point out that only a small fraction (7 out of $10^5$) of trial dynamical simulations of the system remain stable over $\sim$160 Myr.  In addition, the study of the HR 8799 debris disk carried out by \citet{srs09}, indicates the presence of an inner warm ($\sim$150 K) debris belt (see Figure~\ref{contrasts}).  The outer boundary of this disk is most likely sculpted by the ``e'' component at 14 AU. \citet{srs09} suggest that few, if any, dust grains exist interior to 6 AU, and postulate that additional planets may be responsible for the clearing.  Our measurements show no evidence for companions more massive than $\sim$12 M$_{\rm Jup}$ at 3-6 AU, but they do not rule out the possibility that an object less massive than $\sim$12 M$_{\rm Jup}$ was responsible for this inner clearing. Indeed, even relatively low mass objects (0.1-3 M$_{\rm Jup}$) can shepherd debris disk belts \citep[e.g.][]{qbf04,ckk09}.


 
\subsection{Future Work}

Achieving very small inner working angles (10-250 mas) corresponding to projected orbital separations of a few AU will be in crucial for the future ground and space-based characterization of the circumstellar regions around young stars.  Reaching appreciable contrast at such small inner working angles will allow significant sensitivity to objects orbiting much closer to the host star.  Perhaps more importantly, the small inner working angles will allow observers to probe solar system scales of more distant targets, thereby increasing the number of available target stars, including those young stars in the nearest star forming regions \citep{kim08}.   Though traditional high contrast techniques such as dual imaging polarimetry \citep[e.g.][]{obh08} can achieve significant contrasts of $\sim$8-10 mag within 500 mas \citep[e.g.][]{hos09}, the success of these techniques largely depends on a high polarization signal exhibited by the planets.  However, NRM will have superior sensitivity at small inner working angles (20-300 mas) and will play a major role for ground based high contrast imaging of exoplanets and imaging close binaries \citep[e.g.][]{lmi06,hmo11}.  Recent results suggest $\sim$7.5 magnitudes of $K$-band contrast at Keck has  been achieved for $K=8$ star in three hours of observing (A. Kraus, private communication, 2010).

\begin{deluxetable}{lccc}
\tabletypesize{\scriptsize}
\tablecaption{NRM Detection Limits and Mass Upper Limits to Additional HR 8799 companions 
}
\tablewidth{0pt}
\tablehead{ 
\colhead{Sep.~(AU)}  & \colhead{Ang.~Sep.~(mas)} & \colhead{$\Delta L'$ (mag)} & \colhead{Mass Limit (M$_{\rm Jup}$)}\footnotemark[1] }
\startdata
0.8      & 20           & 5.17          & 80  \\
1         & 25           & 5.63           & 60  \\
2         & 50           & 7.48           & 13  \\
3         & 75           & 7.99           & 11  \\
3 - 10 & 75 - 750 & 7.99 - 7.63 & 11 - 12.5
\enddata
\label{contrasttable}
\end{deluxetable}
\footnotetext[1]{Mass limits calculated using the \citet{bcb03} models assuming an age of 30 Myr.}

The science return from NRM will be enhanced when the imaging science camera has multi-wavelength capabilities such as an integral field spectrograph (IFS). This has already been achieved with the Project 1640 IFS and  coronagraph at Palomar Observatory \citep[][Zimmerman et al. 2011, in prep]{hoz11}, and will be integrated with the Gemini Planet Imager coronagraph \citep{mgp08, sso09}. Furthermore, aperture masking will continue to be an efficient tool for characterizing stellar multiplicity.  As several authors have suggested, more massive stellar systems may be more efficient at producing faint stellar companions \citep{dvf09,kmy10}.  Aperture masking will play a crucial role to supplement ongoing surveys of A star multiplicity \citep[e.g.][]{hob10,zoh10}.

In the more distant future, NRM will play a prominent role for the planet-finding efforts of JWST \citep{bkt10}, significantly improving on the sensitivity presented in our study. The Tunable Filter Instrument camera which is integrated with the Fine Guidance System on JWST \citep[e.g.][]{slt10, dhr10} will incorporate non-redundant aperture masking capabilities. Recent simulations indicate a limiting delta magnitude of $\sim$9.5 for integration times of a few hundred seconds in the JWST filters corresponding to the $L'$ and $M$-bands.  These limits will provide sensitivity to objects of a few Jupiter masses for systems at ages of 10-100 Myr \citep{slt10}.


\section{Conclusions}
The results presented in this work are the first constraints on the masses of additional companions to the HR 8799 system at projected orbital separations inside of 10 AU. We can rule out the presence of a very massive perturber that has ejected the four imaged planets out beyond their presumed formation region near the ice line.  Further, for an assumed age of 30 Myr, we place upper limits to planetary mass companions of 80, 60, and 11 M$_{\rm Jup}$ at projected orbital separations of 0.8, 1, and 3 AU, respectively. These findings will be essential for future works to begin clarifying the complicated dynamical interplay of the four known objects in the system, but also furthering our understanding the complicated interaction between these companions and the debris disk belts in the system.  These results may also illuminate future work dedicated to understanding the dust-free hole interior to $\sim$6 AU.



\acknowledgments
This work was performed in part under contract with the California Institute of Technology (Caltech) funded by NASA through the Sagan Fellowship Program. ALK was suported by NASA through Hubble Fellowship grant 51257.01 awarded by the STScI, which is operated by AURA, Inc., for NASA, under contract NAS 5-26555. The data presented herein were obtained at the W.M. Keck Observatory, which is operated as a scientific partnership among the California Institute of Technology, the University of California and NASA. The Observatory was made possible by the generous financial support of the W.M. Keck Foundation.  The authors wish to recognize and acknowledge the very significant cultural role and reverence that the summit of Mauna Kea has always had within the indigenous Hawaiian community.  We are most fortunate to have the opportunity to conduct observations from this mountain.





\end{document}